\documentclass[prl,aps,epsfig,superscriptaddress,showpacs,twocolumn,notitlepage]{revtex4-1}
\usepackage{amsmath,amssymb,amsfonts,amstext}
\usepackage{graphicx,breakurl,color,setspace}
\usepackage[T1]{fontenc}
\usepackage[normalem]{ulem}
\usepackage[unicode=true,pdfusetitle,pdfborder={0 0 1}]{hyperref}
\usepackage{upgreek}
\usepackage{bm}
\usepackage{mathrsfs}
\usepackage{multirow}
\usepackage{tabularx}

\hypersetup{colorlinks,
linkcolor=blue,          citecolor=blue,        filecolor=blue,      urlcolor=blue}

\usepackage{times}

\begin{document}
\title{Observation of non-Hermitian topology with non-unitary dynamics of solid-state spins}
\date{\today}
\author{Wengang Zhang}\thanks{These authors contributed equally to this work.}
\author{Xiaolong Ouyang}\thanks{These authors contributed equally to this work.}
\author{Xianzhi Huang}\author{Xin Wang}\author{Huili Zhang}\author{Yefei Yu}\author{Xiuying Chang}\author{Yanqing Liu}
\affiliation{Center for Quantum Information, IIIS, Tsinghua University, Beijing 100084, P. R. China}

\author{Dong-Ling Deng}\email{dldeng@tsinghua.edu.cn}
\affiliation{Center for Quantum Information, IIIS, Tsinghua University, Beijing 100084, P. R. China}
\affiliation{Shanghai Qi Zhi Institute, 41th Floor, AI Tower, No. 701 Yunjin Road, Xuhui District, Shanghai 200232, China}

\author{L.-M. Duan}\email{lmduan@tsinghua.edu.cn}
\affiliation{Center for Quantum Information, IIIS, Tsinghua University, Beijing 100084, P. R. China}

\begin{abstract}
Non-Hermitian topological phases exhibit a number of exotic features that have no Hermitian counterparts, including the skin effect and breakdown of the  conventional bulk-boundary correspondence.   Here, we implement the non-Hermitian Su-Schrieffer-Heeger (SSH) Hamiltonian, which is a prototypical model for studying non-Hermitian topological phases, with a solid-state quantum simulator consisting of an electron spin and a $^{13}$C nuclear spin in a nitrogen-vacancy (NV) center  in a diamond.  By employing a dilation method, we realize the desired non-unitary dynamics for the electron spin and map out its spin texture in the momentum space, from which the corresponding topological invariant can be obtained directly. Our result paves the way for further exploiting and understanding the intriguing properties of non-Hermitian topological phases with solid-state spins or other quantum simulation platforms.
\end{abstract}

\maketitle

\begin{figure}
	\centering
	\includegraphics[width=\linewidth] {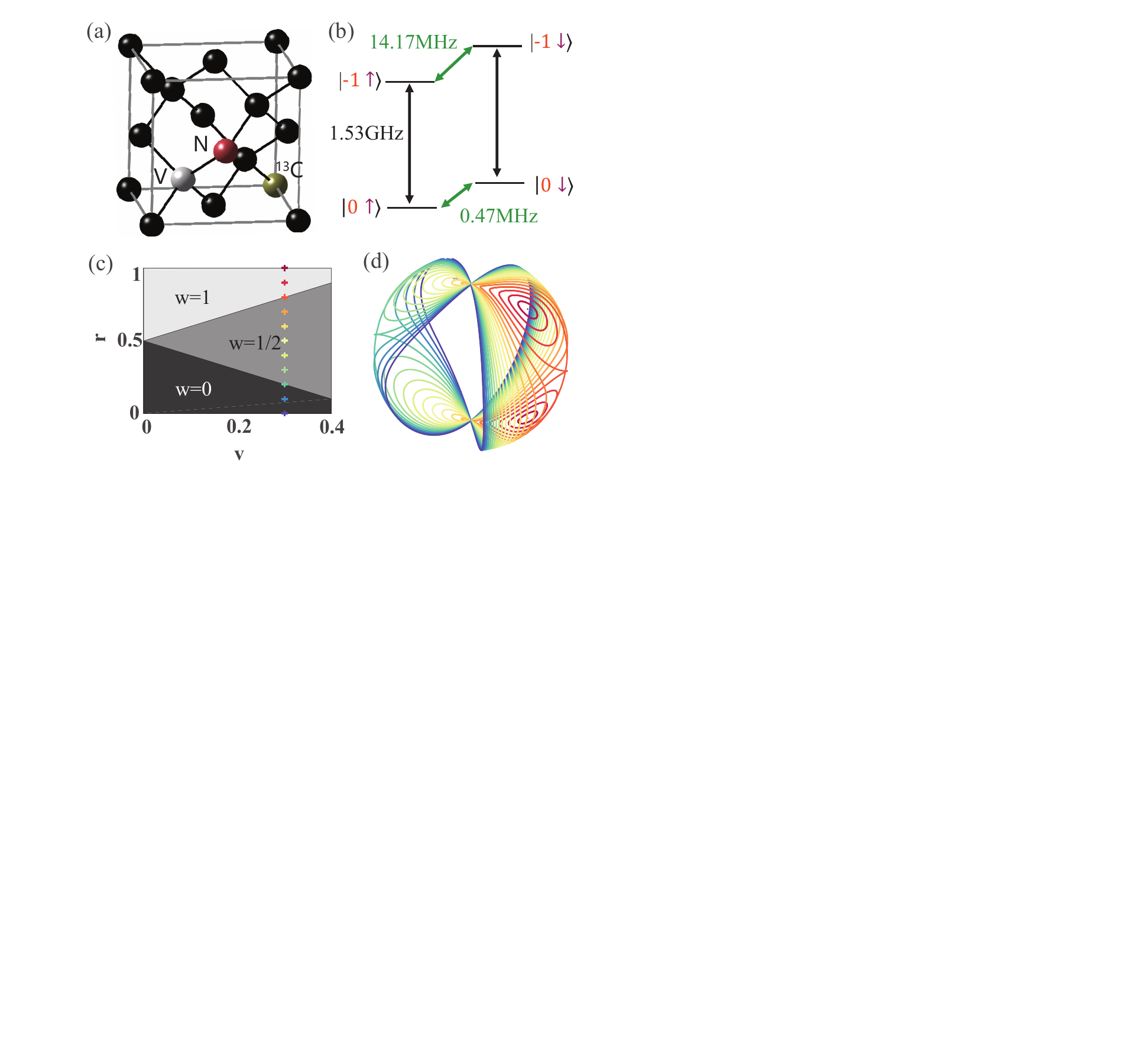}
	\caption{Experimental system and topological properties of the non-Hermitian Su-Schrieffer-Heeger (SSH) model.
		(a), Illustration of the NV center platform used in this work. The electron spin is coupled to a nearby $^{13}$C nuclear spin with $13.7$ MHz hyperfine interaction. (b), Energy level structure of the electron-nuclear spin system. Black arrows indicate the microwave transitions. Green arrows indicate the radio frequency transition. We choose $|0\rangle$ and $|-1\rangle$ to consist our electron-spin qubit. $|\uparrow\rangle$ and $|\downarrow\rangle$ denote the nuclear spin state. (c), A sketch of the phase diagram of the non-Hermitian SSH Hamiltonian $H(k)$. (d), Trajectories of the eigenvectors of $H(k)$ with varying parameters as $k$ sweeps through the Brillouin zone. Different trajectories correspond to the markers with the same color in (c).}
	\label{NVcenter}
\end{figure}

While Hermiticity lies at the heart of quantum mechanics, non-Hermitian Hamiltonians have widespread applications as well \cite{moiseyev2011non,Konotop2016Nonlinear, ashida2020non}. 
Indeed, they have been extensively studied in photonic systems with loss and gain \cite{feng2013experimental, peng2014loss, el2018non, feng2017non, Ozawa2019Topological}, open quantum systems \cite{Dalibard1992Wave, Anglin1997Cold, rotter2009non, zhen2015spawning, diehl2011topology, verstraete2009quantum}, and quasiparticles with finite lifetimes \cite{Shen2018Quantum, zhou2018observation, Yoshida2018Nonhermitian}, etc. More recently, the interplay between non-Hermiticity and topology has attracted tremendous attention \cite{bergholtz2019exceptional,coulais2020topology}, giving rise to an emergent research frontier of non-Hermitian topological phases of matter. In contrast to topological phases for Hermitian systems \cite{Qi2011Topological,Hasan2010Colloquium,Chiu2016Classification}, non-Hermitian ones bears several peculiar features, such as the skin effect \cite{McDonald2018Phase,Kunst2018Biorthogonal,Yao2018Edge} and breakdown of the conventional bulk-boundary correspondence \cite{Kunst2018Biorthogonal,Yao2018Edge,Lee,Yokomizo2019NonBloch,alvarez2018topological,Borgnia2020NonHermitian},  
and new topological classifications \cite{Kawabata2019Symmetry, Zhou2019Periodic,kawabata2019topological}.  Experimental observations of the non-Hermitian skin effect  have been reported in mechanical metamaterials \cite{ghatak2019observation}, non-reciprocal topolectric circuits \cite{helbig2019observation}, and photonic systems \cite{xiao2020non,Zhu2020Photonic,weidemann2020topological}.  However, despite the notable progress, direct observation of the topological invariant for non-Hermitian systems has not been reported in a quantum solid-state system hitherto, owing to the stringent requirement of delicate engineering of the coupling between the target system and the environment in implementing non-Hermitian Hamiltonians.  In this paper, we carry out such an experiment and report the direct observation of non-Hermitian topological invariant with a solid-state quantum simulator consisting of both electron and  nuclear spins in a NV center (see Fig. \ref{NVcenter}).

NV centers in diamonds \cite{doherty2013nitrogen} exhibit atom-like properties, such as long-lived spin quantum states and well-defined optical transitions, which make them an excellent experimental platform for quantum information processing \cite{wrachtrup2001quantum,wu2019programmable,bernien2013heralded,humphreys2018deterministic,schmitt2017submillihertz,zhou2014quantum}, sensing \cite{Balasubramanian2008Nanoscale,kolkowitz2015probing,kucsko2013nanometre}, and quantum simulation \cite{yuan2017observation,lian2019machine,kucsko2018critical}.  For Hermitian topological phases, simulations of three dimensional (3D) Hopf insulators \cite{yuan2017observation} and chrial topological insulators \cite{lian2019machine} with NV centers have been demonstrated in recent experiments, and observations of their topological properties, such as nontrivial topological links associated with the Hopf fibration and the integer-valued topological invariants, have been reported. A key idea that enables these simulations is to use the adiabatic passage technique, where we treat the momentum-space  Hamiltonian as a time-dependent one with the momentum playing the role of time. The ground state of the Hamiltonian at different momentum points can be obtained via adiabatically tuning the frequency and the amplitude of a microwave that manipulates the electron spin in the center, and quantum tomography of the final state with a varying momentum provides all the information needed for obtaining the characteristic topological properties \cite{yuan2017observation}.

Nevertheless, simulating non-Hermitian topological phases with the NV center platform (see Fig.\ref{NVcenter} (a, b)) faces two
apparent challenges. First, for non-Hermitian systems the governing Hamiltonians typically have complex eigenenergies and the conventional adiabatic theorem is not necessarily valid in general \cite{lbanez2014Adiabaticity}. As a consequence, the adiabatic passage technique does not apply and the preparation of eigenstates of non-Hermitian Hamiltonians becomes trickier. Second, the non-Hermiticity requires a
delicate engineering of the coupling between the targeted system and the environment, so that tracing out the environment could leave the system effectively governed by a given non-Hermitian Hamiltonian. These two challenges make simulating non-Hermitian topological phases notably more difficult than that for their Hermitian counterparts.
In this paper, we overcome these two challenges and report the first experimental demonstration of simulating non-Hermitian topological phases with the NV center platform. In particular, we implement a prototypical model for studying non-Hermitian topological phases, i.e., the non-Hermitian SSH model, by carefully engineering the coupling between the electron and nuclear spins through a dilation method that was recently introduced for studying parity-time symmetry breaking with NV centers \cite{wu2019observation}. Without using adiabatic passage, we find that the non-unitary dynamics generated by the non-Hermitian Hamiltonian will autonomously drive the electron spin into the eigenstate of the Hamiltonian with the largest imaginary eigenvalue, independent of its initial state. The topological nature of the Hamiltonian can be visualized  by mapping out the spin texture in the momentum space and the topological invariant can be derived directly by a discretized integration over the momentum space.

We consider the following non-Hermitian SSH model Hamiltonian in the momentum space \cite{Yao2018Edge,Lee}:
\begin{eqnarray}
H(k)=\gamma [h_x\sigma_x+(h_z+\frac{i}{2})\sigma_z],
\end{eqnarray}
where $\gamma$ measures the energy scale (we set $\hbar=1$ for simplicity),  $h_x=v+r\cos k$, $h_z=r\sin k$, and $\sigma_{x,z}$ are the usual Pauli matrices. This Hamiltonian possesses a chiral symmetry $\sigma_y^{-1}H(k)\sigma_y=-H(k)$, which ensures that its eigenvalues  appear in $(E,-E)$ pairs. Its energy gap closes at the exceptional points $(h_x,h_z)=(\pm \frac{1}{2}, 0)$, which gives $v=r\pm \frac{1}{2}$ for $k=\pi$ and $v=-r\pm \frac{1}{2}$ for $k=0$. The topological properties for the Hamiltonian can be characterized by the winding number $w$ of
$H(k)$, circling around the exceptional points as $k$ sweeps through the first Brillouin zone \cite{Lee}: $w=0$, $\frac{1}{2}$, and $1$ respectively, if $H(k)$ encircles zero, one, and two exceptional points.  A sketch of the phase diagram of the non-Hermitian SSH model is shown in Fig. \ref{NVcenter} {\bf c}.


To implement the non-Hermitian Hamiltonian $H(k)$ with the NV center platform, we exploit a dilation method introduced in Ref. \cite{wu2019observation}. We use the electron spin as the targeted system and a nearby $^{13}$C nuclear spin as the ancillary qubit. Suppose that the dynamics of the electron spin is described by $H_e$ and the dilated system described by $H_\text{e,n}$, then the problem essentially  reduces to a task that for a given momentum $k$ we need to carefully engineer $H_\text{e,n}$,  such that $H_e$ equals $H(k)$ after projecting  the nuclear spin onto a desired state. The basic idea is as follows.
We consider a quantum state $|\psi\rangle$ evolving under a non-Hermitian Hamiltonian $H_e$, which satisfies the Schr{\"o}dinger equation $i  \frac{\partial }{\partial t} |\psi(t)\rangle= H_e |\psi(t)\rangle$. Then we introduce a dilated state $|\Psi(t)\rangle = |\psi(t)\rangle|-\rangle+\eta(t)|\psi(t)\rangle |+\rangle $ governed by the dilated Hermitian Hamiltonian $H_\text{e,n}$. Here, $|-\rangle = (|\uparrow\rangle-i|\downarrow\rangle)/\sqrt{2}$ , $|+\rangle = -i(|\uparrow\rangle+i|\downarrow\rangle)/\sqrt{2}$ and $\eta(t)$ is a proper time-dependent linear operator. The dilated system satisfies $i  \frac{\partial }{\partial t} |\Psi(t)\rangle= H_\text{e,n} |\Psi(t)\rangle$. For our purpose, the dilated Hamiltonian should be designed properly as (see Supplementary Information):
\begin{eqnarray}
H_{\text{e,n}}&=&[A_0(t)I +A_1(t)\sigma_x +A_2(t)\sigma_y  +A_3(t)\sigma_z]\otimes \mathbf{I} \nonumber \\
&+& [B_0(t)I+B_1(t)\sigma_x+B_2(t)\sigma_y+B_3(t)\sigma_z]\otimes\sigma_z, \nonumber
\end{eqnarray}
where $\mathbf{I}$ is the two-by-two identity matrix, and $A_i(t)$ and $B_i(t)$ ($i=0,1,2,3$) are time-dependent real-valued functions determined by $H_e$. After the time evolution process, we can project the nuclear spin onto its $|-\rangle\langle -|$ subspace to obtain $|\psi(t)\rangle$.  

\begin{figure}
	\centering
	\includegraphics[width=1\linewidth] {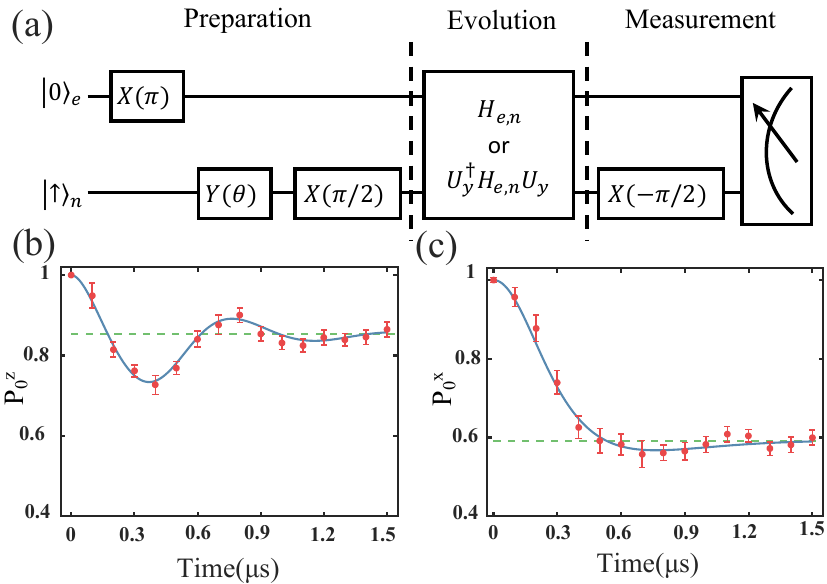}
	\caption{The quantum circuits and the benchmark of the non-unitary time-evolution. (a), The quantum circuits used in our experiment. Through optical pumping, we first polarize the electron and nuclear spins onto $|0\rangle_e$ and $|\uparrow\rangle_n$, respectively.  Then, rotations along x- and y-axises will prepare the dilated system onto the state $|\Psi(0)\rangle=|-1\rangle_e|-\rangle_n+\eta(0)|-1\rangle_e|+\rangle_n$. The evolution box implements the unitary dynamics generated by the dilated Hamiltonian $H_\text{e,n}$, after which we measure the nuclear spin in the $|\pm\rangle$ basis. A postselection of nuclear spin in the $|-\rangle$ state collapses the electron spin into the desired eigenstate of $H_e=H(k)$ for a given momentum $k$. (b,\;c), The time evolution of the populations $P^z_0=\text{Tr}(\rho_e|0\rangle_e\langle0|)$ and  $P^{x}_0=\text{Tr}(\rho_e|+\rangle_e\langle +|)$ with $|+\rangle_e=\frac{1}{\sqrt{2}}(|0\rangle+|-1\rangle)$.  For (b) and (c), the parameters characterizing the underlying Hamiltonian are chosen as $v=0.3$, $r=1$, $\gamma=3.5$, and $k=0.3\pi$, and $v=0.3$, $r=0.3$, $\gamma=4$, and $k=0.6\pi$, respectively. Here, the solid lines plot the results from numerical simulations and  the dashed lines indicate the ideal population of the targeted state at the long time limit. }
	\label{sequence}
\end{figure}

\begin{figure*}
	\centering
	\includegraphics[width=0.9\linewidth] {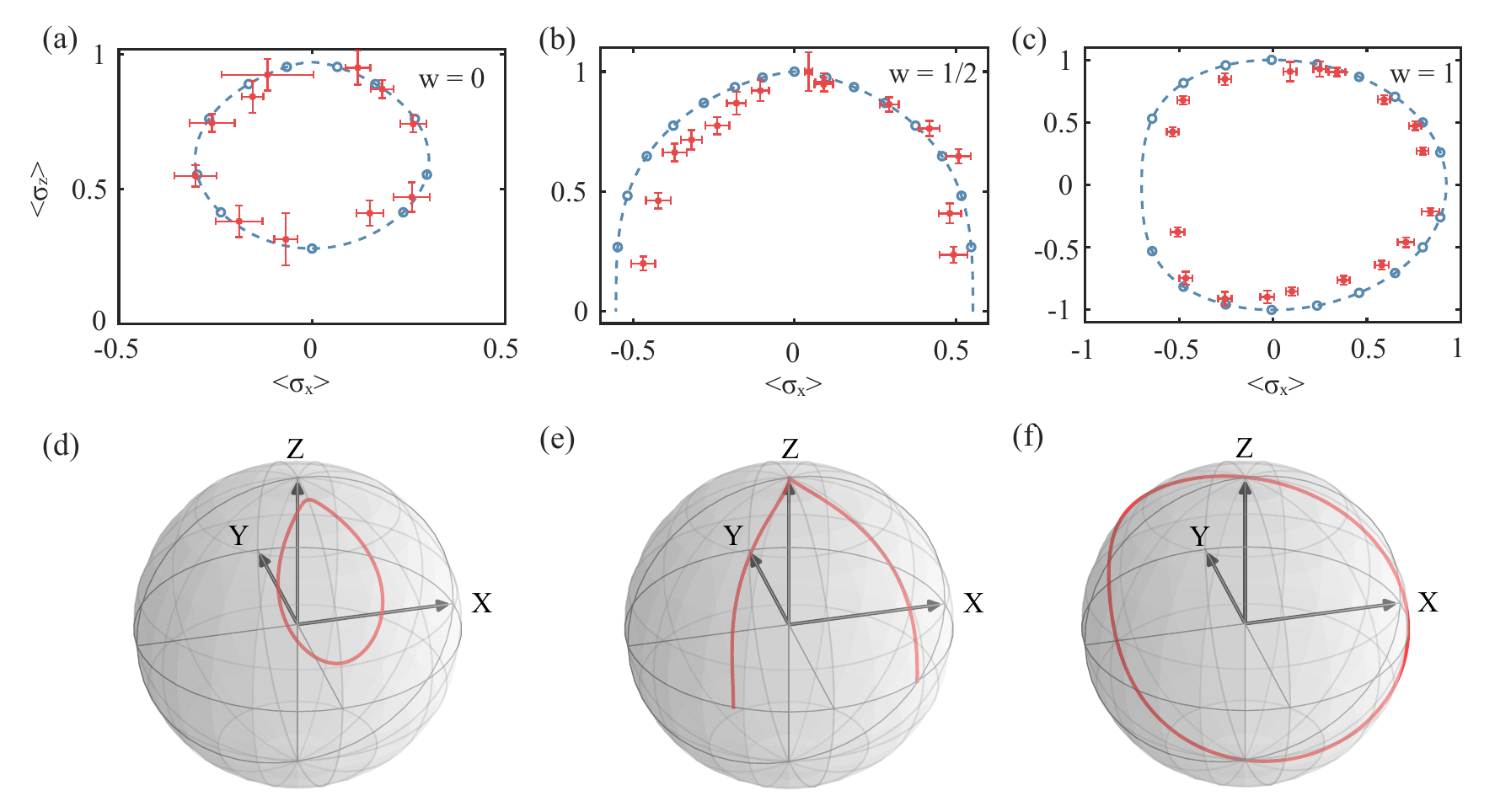}
	\caption{The trajectory of the target eigenvector of $H(k)$ in the Brillouin zone. (a), (b), and (c) plot the experimental results of $\langle\sigma_z\rangle$ versus $\langle\sigma_x\rangle$  as $k$ sweeps the first Brillouin zone. Here, the red dots with error bars denote the experimental data, whereas the blue dots with dashed curves represent the theoretical prediction from numerical simulations. The parameters are chosen as $v=0.3$ and $r=0.18,0.3,1$ in (a), (b), and (c) respectively, with corresponding winding number $w=0,\frac{1}{2},1$. (d), (e), and (f) show the theoretical trajectories of the electron spin state on the Bloch sphere, whose projection onto the xz-plane correspond to (a), (b) and (c) respectively.
	}
	\label{topodata}
\end{figure*}

Unlike the case of simulating Hermitian topological phases \cite{yuan2017observation,lian2019machine}, where the ground state of the Hamiltonian at different momentum points can be obtained through adiabatic passages, for non-Hermitian Hamiltonians their eigenvalues are complex numbers in general and the adiabatic passage method does not apply. Fortunately, we can explore the non-unitary dynamics generated by the non-Hermitian Hamiltonian to prepare the eigenstate  that corresponds to the eigenvalue with the largest imaginary part. To be more specific, suppose the electron spin is initially at an arbitrary state $|\psi(0)\rangle=\alpha_1 |R_1\rangle +\alpha_2|R_2\rangle$, where $|R_{1,2}\rangle$ are the right eigenstates of $H_e$ corresponding to the eigenvalues $\lambda_{1,2}$.  Without loss of generality, we assume $\text{Im}(\lambda_1)>\text{Im}(\lambda_2)$. Then the electron spin state will decay to $|R_1\rangle$ in the long time limit.
As a result, we can prepare the eigenstate $|R_1\rangle$ of $H_e$ by just waiting long enough time for the system to decay to this state.

To experimentally realize $H_\text{e,n}$, we apply two microwave pulses with time-dependent amplitude, frequency and phase. We explore the state evolution by monitoring the population on $|0\rangle_e$ state and see how it decays to the desired eigenstate of $H_e$. In Fig. \ref{sequence} (a), we show the quantum circuits used in our experiment.  Fig. \ref{sequence} (b) and (c) show our experimental results of $P_0^{z,x}$ as a function of time. From these figures, it is evident that our experimental results match the theoretical predictions excellently, within the error bars for almost all of the data points. In addition, after long enough evolution time (about $1.5\;\mu s$ in our experiment), the electron spin state decays to the desired eigenstate of $H(k)$ for different momentum $k$. This indicates that our dilated Hamiltonian indeed effectively implement the non-Hermitian $H(k)$ in our experiment.

We mention that in Fig. \ref{sequence}(c), we measure $P_0$ in the x-basis, which requires  an
 $\pi/2$ rotation of the electron spin.
 In order to avoid off-resonance driving, the microwave driving power should be weak enough. The Rabi frequency should be much smaller than the hyperfine coupling strength ($13.7$ MHz), so that the $\pi/2$ rotation would take time on the order of a microsecond.  Meanwhile, the time evolution process also takes approximately half of the system's coherence time ($T_2^\star=3.3\; \mu s$).  Thus, adding an additional rotation would not only increase the processing time but also introduce both gate and decoherence errors. To avoid this, we first apply a unitary transform to the target Hamiltonian: $\widetilde{H_e}=U_y^\dagger H_eU_y$, where $U_y=\frac{1}{\sqrt{2}}\begin{pmatrix} 1&-1 \\1&1 \end{pmatrix}$. We evolve the electron spin with $\widetilde{H_e}$ instead of $H_e$ and measure the final state in the z-basis. This is equivalent to evolving the electron spin with $H_e$ and then measuring in the x-basis, but with improved efficiency and accuracy (see the Supplementary Information).  In addition, the time needed for the initial state to decay to the desired eigenstate of $H_e$ depends crucially on the difference between the imaginary parts of its two eigenvalues. For some parameter regions, this difference may  not be large enough and the decay time  could even be longer than $T_2^\star$. In order to speedup the process, we increase $\gamma$ according to the specific parameters so that we can finish the experiment within the coherence time.

To probe the topological properties of the non-Hermitian SSH model, we can measure $\langle\sigma_z\rangle$ and $\langle\sigma_x\rangle$ for the final state of the time evolution [which is basically the eigenstate $|R_1\rangle$ of $H(k)$] as $k$ sweeps the first Brillouin zone. We plot our experimental results in Fig. \ref{topodata}. When $H(k)$ encircles no or two  exceptional points, the eigenvector of $H(k)$ are $2\pi$ periodic in $k$, and the trajectory of $\langle\sigma_z\rangle$ and $\langle\sigma_x\rangle$  as $k$ sweeps through the Brillouin zone forms closed circles.  In this case, the winding number is zero or one, depending on whether the trajectory of $\langle\sigma_z\rangle$ and $\langle\sigma_x\rangle$ winds around the origin or not, as clearly shown in Fig. \ref{topodata}(a) and (c). In contrast, when $H(k)$ encircles only one exceptional point, the eigenvector will have a $4\pi$ periodicity and $k$ must sweep through $4\pi$ to close the trajectory, giving rise to a fractional value of the winding number $w=\frac{1}{2}$ if $k$ only sweeps through the first Brillouin zone  \cite{Lee}. This is also explicitly observed in our experiment as shown in Fig. \ref{topodata}(b). We mention that in Fig. \ref{topodata}(c), when $k$ sweeps across $\pi$, the imaginary part of the eigenvalues of $H(k)$ will exchange their sign, leading to a leap from one eigenstate to another. Mathematically, we can prove that $\langle R_1|\sigma_{z,x}|R_1\rangle=-\langle R_2|\sigma_{z,x}|R_2\rangle$  and $\langle R_1|\sigma_y|R_1\rangle=\langle R_2|\sigma_y|R_2\rangle$ (see Supplementary Information). As a result, in the experiment we obtain $\langle R_1|\sigma_{z,x}|R_1\rangle$ by actually measuring $\langle R_2|\sigma_{z,x}|R_2\rangle$ after the crossing of the eigenstates. In addition, with our experimentally measured data the winding number can also be calculated directly through a discretized integration over the momentum space (see Supplementary Information). In Table \ref{TableWindingNumber}, we show the winding number calculated from the experimental and theoretically simulated data for different parameter values of $r$. From this table, it is clear that the winding number calculated from the experimental data matches its theoretical predictions within a good precision,  and is in agreement with that obtained from the trajectory of $\langle\sigma_z\rangle$ and $\langle\sigma_x\rangle$ as well. 

\begin{table}
	\centering
	\includegraphics[width=\linewidth] {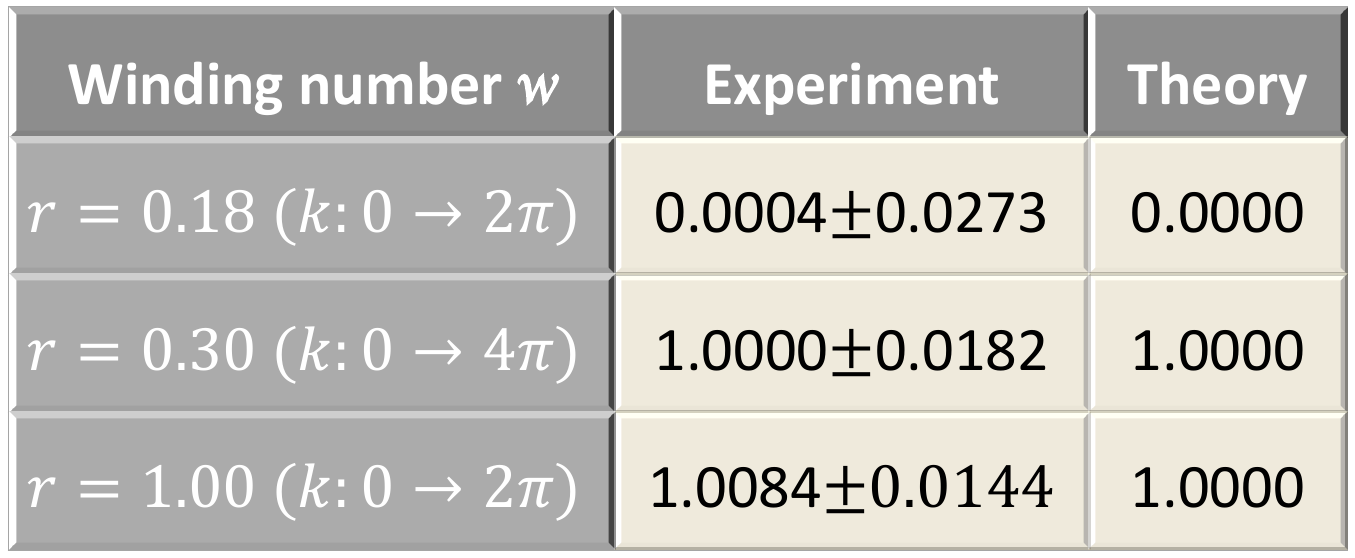}
	\caption{The winding number extracted from the experimental and numerically simulated data. Here, the model parameters are chosen the same as in Fig. \ref{topodata}. For $r=0.3$, $H(k)$ only encircles one exceptional point and its eigenvector has a $4\pi$ periodicity. As a result, for this case we sweep the momentum $k$ through $[0,4\pi]$ to form a closed loop so that the discretized integration for computing the winding number is well-defined (see Supplementary Information). 
		}
	\label{TableWindingNumber}
\end{table}

In summary, we have experimentally observed the non-Hermitian topological properties of the SSH model  through non-unitary dynamics with a solid state quantum simulator. Our method carries over straightforwardly to other types of non-Hermitian topological models that are predicted to exist in the extended periodic table but have not yet been observed in any experiment. It thus paves the way for future explorations of exotic non-Hermitian topological phases with the NV center or other quantum simulation platforms.


We acknowledge helpful discussions with Yong Xu,  Yukai Wu and Liwei Yu.
This work was supported by the Frontier Science Center for Quantum Information of the Ministry of Education of China, Tsinghua University Initiative Scientific Research Program, the Beijing Academy of Quantum Information Sciences, and the National key Research and Development Program of China (2016YFA0301902). D.-L. D. also acknowledges additional support from the Shanghai Qi Zhi Institute.

\section{Supplementary Information: Observation of non-Hermitian topology with non-unitary dynamics of solid-state spins}

\section{Experimental setup}
Our sample is mounted on a laser confocal system. A $532$ nm green laser is used for off-resonance excitation. The laser pulses are modulated by an acousto-optic modulator (AOM). We use an oil objective lens to focus the laser beam onto the damond sample. In addition, this lens is also used to collect the fluorescence photons.  The fluorescence photons are detected by a single photon detector module (SPDM) and counted by a homemade field-programmable gate array (FPGA) board. The $480$ Gauss magnetic field is provided by a permanent magnet along the NV axis.

An arbitrary waveform generator (AWG, Techtronix 5014C) is used to generate low frequency analog signals and transistor-transistor logic (TTL) signals. Two of the analog signals are used to modulate the carrier microwave (MW) signal (generated by a MW source, Keysight N5181B) through an IQ-mixer so as to control its  phase, amplitude and frequency conveniently. Another analog signal of AWG is used for generating the radio frequency (RF) signal. The TTL signals are used for modulating the laser pulse and providing gate signals for the FPGA board. The MW signal is applied onto the sample through a homemade MW coplanar waveguide. The RF signal is applied through a homemade coil. Before applied onto the sample, both MW and RF signals are amplified via amplifiers (Mini Circuits ZHL-30W-252-S+ for MW and Mini Circuits LZY-22+ for RF).

\section{The sample}
\begin{figure}[h]
	\centering
	\includegraphics[width=1\linewidth] {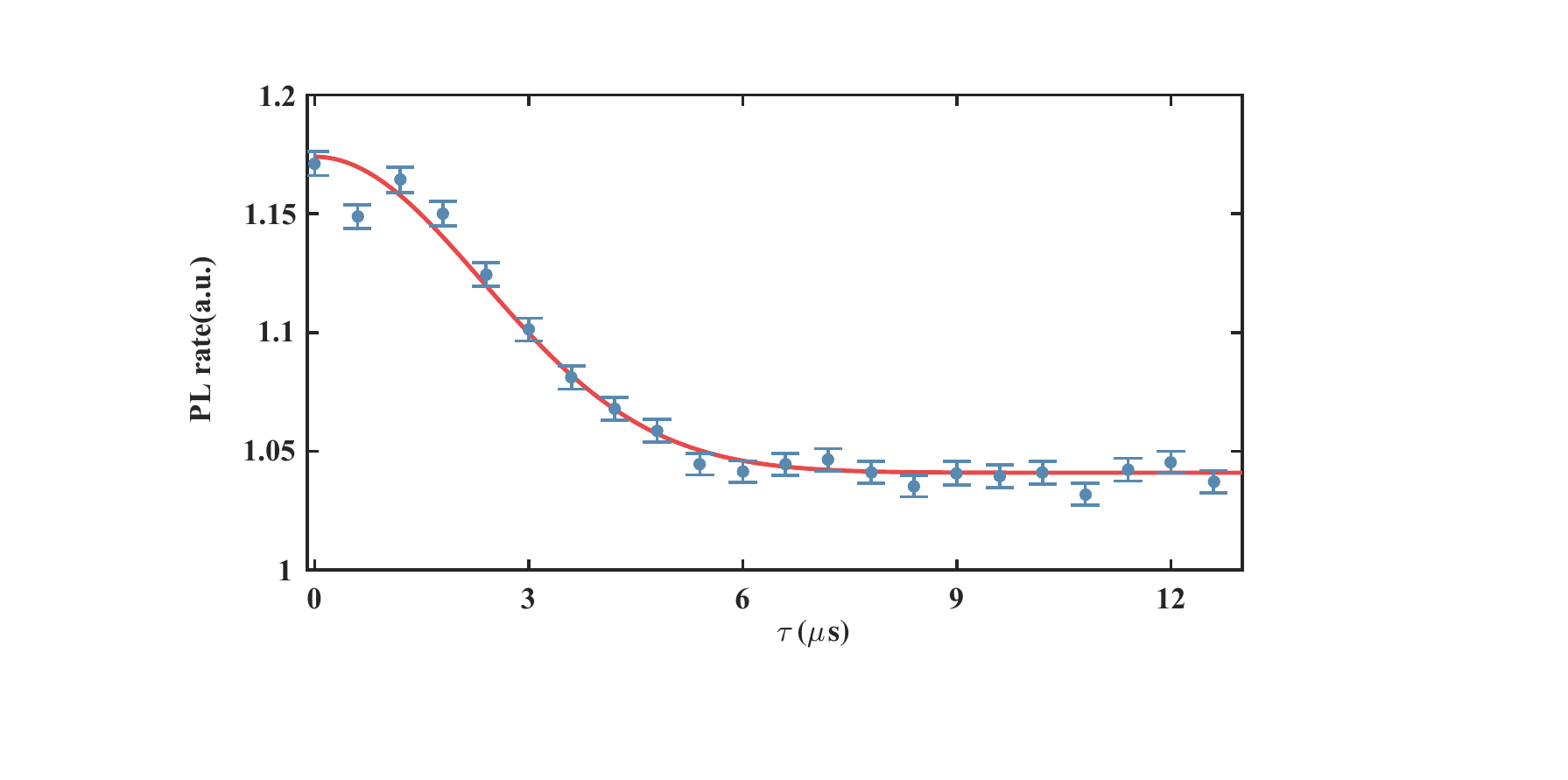}
	\caption{\textbf{Dephasing time measurement.}
		We use the Ramsey experiment to verify the dephasing time of the electron spin. Solid line is the fit to the experimental data. The $T_2^*$ is measured to be $3.3 \mu s$.
	}
	\label{ramsey}
\end{figure}

Our experiment is performed on an electronic grade diamond produced by Element Six with a natural abundance of $1.1\%$ for $^{13}$C. The crystal orientation is along the $\langle 100 \rangle$ direction. The electron spin in the NV center is used as the target qubit and a nearby $^{13}$C nuclear spin is used as the ancilla qubit. 
We use the Ramsey experiment to obtain the $T_2^*$ of the electron spin (see Fig. \ref{ramsey}). For the sample used in our experiment, the $T_2^*$ is measured to be $3.3 \mu s$. Since the time evolution process employed in our experiment is no longer than  $1.8\mu s$, so the coherence time is long enough for our purpose. 

\section{More details about the NV center}
\subsection{Spin initialization}
In this section, we briefly introduce the spin initialization process in our experiment. A 532nm green laser can be used to off-resonantly excite the NV center. Because of the intersystem crossing (ISC) \cite{goldman2015state}, this process can be used to initialize the electron spin state onto the $|m_s=0\rangle$ state. The initialization fidelity is estimated to be $89\%$ \cite{robledo2011high}. Meanwhile, when we apply a magnetic field around $500$ Gauss (which is $480$ Gauss in our experiment), there exists a flip-flop process between electron and nuclear spins  due to the excited-state level anticrossing (ESLAC) \cite{Jacques2009Dynamic}, when the electron spin is in the excited state. Thus, the optical pumping process will polarize not only  the electron spin, but also the nuclear spin  \cite{smeltzer2009robust}.

\subsection{The system Hamiltonian}
The NV center has a spin-triplet ground state. Together with a strongly coupled $^{13}$C nuclear spin, it forms a highly controllable two-qubit system. By applying an external magnetic field along the quantization axis, the Hamiltonian of the NV center can be written as:
\begin{equation}
H_{\text{NV}}= DS_z^2+\omega_e S_z + \omega_n I_z + A_\text{zz}S_zI_z, \label{NVHam}
\end{equation}
where we use secular approximation to dump all terms which are not commute with $S_z$. Here, $D=2\pi\times 2.87$GHz is the zero-field splitting of electron spin; $\omega_e=\gamma_e B$ ($\omega_n=\gamma_n B$) is the Zeeman splitting of the electron (nuclear) spin; $A_\text{zz}=2\pi\times 13.7$MHz is the hyperfine coupling strength.
In our experiment, we use only two level states $|0\rangle$ and $|-1\rangle$ of the electron spin, as shown in Fig. 1\text{b} in the main text. The Hilbert dimension of the dilated system is four, whose basis vectors are denoted as $|0_e,\uparrow_n\rangle$, $|0_e,\downarrow_n\rangle$, $|-1_e,\uparrow_n\rangle$ and $|-1_e,\downarrow_n\rangle$. In this subspace, the Hamiltonian in Eq. (\ref{NVHam})  reduces to:
\begin{equation}
H_0=-(D-\omega_e-\frac{A_\text{zz}}{2})\sigma_z \otimes \mathbf{I}+(\omega_n-\frac{A_\text{zz}}{2})\textbf{I} \otimes \sigma_z+\frac{A_\text{zz}}{2} \sigma_z\otimes \sigma_z.
\label{NV_Hamiltonian}
\end{equation}

\subsection{Spin state readout}
We can readout the spin state by the spin-dependent photoluminescence (PL) rate \cite{zu2014experimental}. The existence of ISC will lead to a decrease of fluorescence rate when electron is in its $|-1\rangle$ state. Meanwhile, due to the ESLAC, different nuclear spin states will also cause a difference on the fluorescence rate. For convenience, we label the states $|0_e,\uparrow_n\rangle$, $|0_e,\downarrow_n\rangle$, $|-1_e,\uparrow_n\rangle$, and $|-1_e,\downarrow_n\rangle$  with numbers from 1 to 4, and denotes their corresponding populations as $P_i$ with $i=1,2,3,4$. By optical pumping, we initialize the system onto $|0_e,\uparrow_n\rangle$ state. By flipping the population onto different states, we can get the PL rates of each different state ($N_i$,$i=1,2,3,4$).

After the time evolution process, the state of our system is $|\Psi(t)\rangle = |\psi(t)_e\rangle|-_n\rangle+\eta(t)|\psi(t)_e\rangle |+_n\rangle $. We apply a nuclear-spin $\pi/2$ rotation to rotate the state into $|\Psi(t)\rangle = |\psi(t)_e\rangle|\uparrow_n\rangle+\eta(t)|\psi(t)_e\rangle |\downarrow_n\rangle $. Here what we want is the expectation value of $\sigma_z$ for the state $|\psi(t)\rangle$. This can be achieved by the renormalized population of state $|0_e,\uparrow_n\rangle$ in the $|\uparrow_n\rangle$ subspace (i.e. $\frac{P_1}{P_1+P_3}$). Hence, we need to know all the populations on the four energy levels. The PL rate of the final state can be described as
\begin{equation}
N_f^0=\Sigma_i P_i N_i.
\end{equation}

After the time evolution process,by flipping the populations between different states, and measure the corresponding PL rates \cite{van2012decoherence}, we can solve the following linear equations to obtain $P_i$s: 
\begin{equation}
\begin{bmatrix}
N_1&N_2&N_3&N_4\\
N_1&N_4&N_3&N_2\\
N_3&N_2&N_1&N_4\\
N_4&N_2&N_1&N_3
\end{bmatrix}
\begin{bmatrix}
P_1\\P_2\\P_3\\P_4
\end{bmatrix}
=
\begin{bmatrix}
N_f^0\\N_f^{\pi_{24}}\\N_f^{\pi_{13}}\\N_f^{\pi_{13}\pi_{34}}
\end{bmatrix}.
\end{equation}
Here, $\pi_{ij}$ represents the $\pi$-pulse between state $i$ and state$j$; $N_f^{\pi_{13}\pi_{34}}$ means that before measuring the PL rate, we apply $\pi_{34}$ and $\pi_{13}$ to flip the population sequentially. 
We use the maximum likelihood estimation \cite{hayashi2005asymptotic} method to reconstruct the final population $P_i$ ($i=1,2,3,4$) under the normalization constraint $P_1+P_2+P_3+P_4=1$.

\section{Construction of the dilated Hamiltonian in the NV center}
\subsection{Construct the dilated Hamiltonian}
In this section, we give more details on how to implement the non-Hermitian SSH model, which is crucial in studying its topological properties. Suppose we have a state $|\psi\rangle$, which evolves under a non-Hermitian Hamiltonian $H_e$ described by the following Schr{\"o}dinger equation (we set $\hbar=1$ here):
\begin{equation}
i \frac{\partial }{\partial t} |\psi(t)\rangle= H_e |\psi(t)\rangle.
\end{equation}

Now we want to find a dilated Hamiltonian $H_{e,n}$, and a dilated system  $|\Psi(t)\rangle = |\psi(t)\rangle|-\rangle+\eta(t)|\psi(t)\rangle |+\rangle $ satisfying
\begin{equation}
i \frac{\partial }{\partial t} |\Psi(t)\rangle= H_{\text{e,n}} |\Psi(t)\rangle,
\end{equation}
where $|-\rangle = (|\uparrow\rangle-i|\downarrow\rangle)/\sqrt{2}$, and $|+\rangle = -i(|\uparrow\rangle+i|\downarrow\rangle)/\sqrt{2}$.

We take the dilated Hamiltonian the form mentioned in Ref \cite{wu2019observation}, such that
\begin{equation}
H_{\text{e,n}}=\Lambda(t)\otimes \mathbf{I} +\Gamma(t)\otimes\sigma_z,
\label{DH}
\end{equation}
where
\begin{eqnarray}
\Lambda(t)&=&\{H_e+[i\frac{d}{dt} \eta(t)+\eta(t)H_e]\eta(t)\}M^{-1}(t), \\
\Gamma(t)&=& i[H_e\eta(t)-\eta(t))H_e-i\frac{d}{dt} \eta(t)]M^{-1}(t), \\
M(t)&=& \eta^\dagger(t)\eta(t)+\mathbf{I}.
\end{eqnarray}

Here the time-dependent operator $M(t)$ satisfies
\begin{equation}
i\frac{d}{dt}M(t)=H_e^\dagger M(t)-M(t)H_e.
\end{equation}
Also, we should choose $M(0)$ properly to make sure $M(t)-\mathbf{I}$ remains positive throughout the experiment.

\subsection{Realize the dilated Hamiltonian in NV center system}
\begin{figure}[h]
	\centering
	\includegraphics[width=1\linewidth] {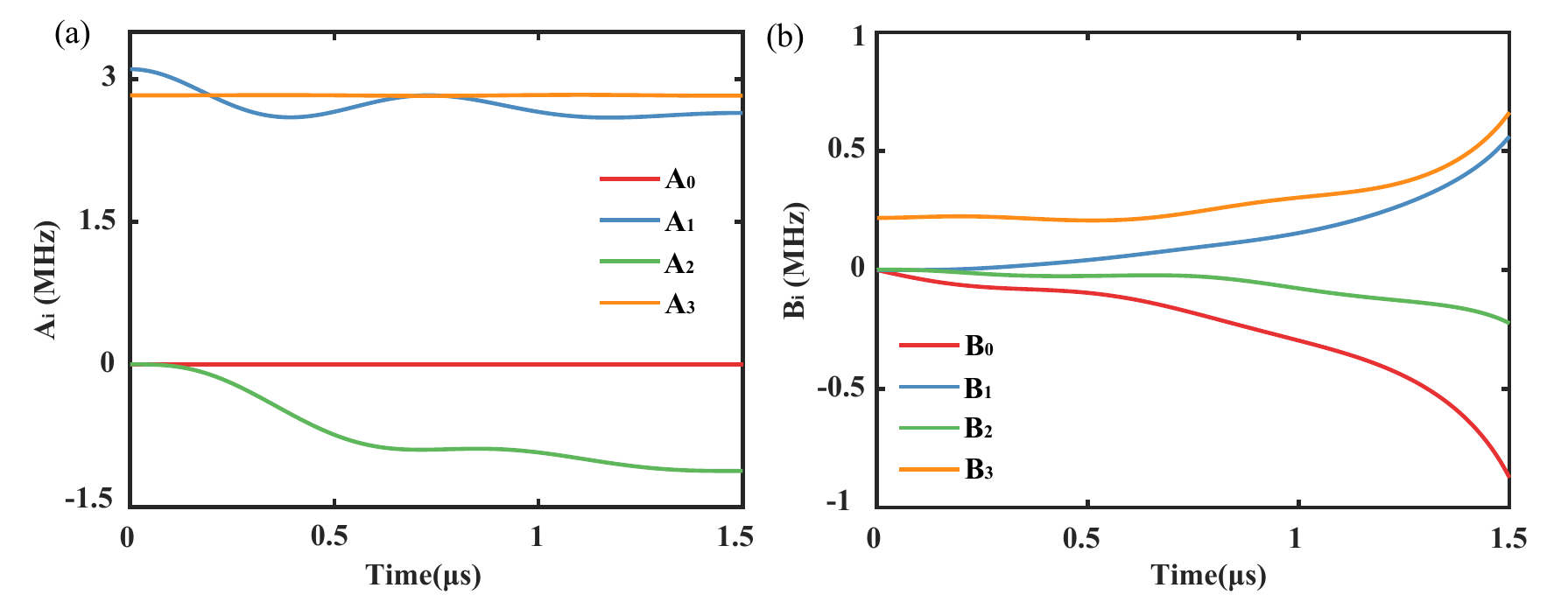}
	\caption{\textbf{Parameters $A_i$ and $B_i$ in the dilated Hamiltonian}. 
		(a) Parameters $A_i$ as functions of time. (b) Parameters $B_i$ as functions of time.}
	\label{AB_parameter}
	
\end{figure}

In this subsection, we introduce how to realize the Hamiltonian $H_\text{e,n}$ in our NV-center system.
To begin with, we expand the Hamiltonian in terms of Pauli operators
\begin{equation}
\begin{split}
H_{\text{e,n}}&=A_0(t)\mathbf{I} \otimes \mathbf{I} +A_1(t)\sigma_x\otimes \mathbf{I}\\
&+A_2(t)\sigma_y\otimes \mathbf{I}+A_3(t)\sigma_z\otimes \mathbf{I} \\ 
&+B_0(t)\mathbf{I}\otimes\sigma_z+B_1(t)\sigma_x\otimes\sigma_z\\
&+B_2(t)\sigma_y\otimes\sigma_z+B_3(t)\sigma_z\otimes\sigma_z,
\end{split}
\label{hsa}
\end{equation}
where $A_i(t)$ and $B_i (t)$ ($i=0,1,2,3$) are time-dependent real parameters. In Fig. \ref{AB_parameter}, we show the time dependence of $A_i(t)$ and $B_i (t)$, with parameters set as $\eta(0)=8, \gamma=3.5, v=0.3$, and $r=1$. 

The Hamiltonian of our NV system is described in Eq. \ref{NV_Hamiltonian}.
Now we apply two individual MW pulses to selectively drive the two different electron spin transitions that are shown in Fig.1\text {b} in the main text. The Hamiltonian induced by the  MW pulse can be written as:
\begin{equation}
\begin{split}
H_{\text{MW}}=\pi*\left(\Omega_1(t)\cos [ \int_{0}^{t}\omega_1(\tau)d\tau +\phi_1(t)] \sigma_x \otimes \mathbf{I} \right.\\
\left. +\Omega_2(t)\cos [ \int_{0}^{t}\omega_2(\tau)d\tau +\phi_2(t)] \sigma_x \otimes  \mathbf{I}\right).
\end{split}
\end{equation}
By choosing the rotating frame
\begin{equation}
U_{\text{rot}}=e^{i\int_{0}^{t}[H_0-A_3(\tau)\sigma_z \otimes  \mathbf{I}-B_0(\tau) \mathbf{I}\otimes\sigma_z-B_3(\tau)\sigma_z\otimes\sigma_z]d\tau}.
\end{equation}
we have the effective Hamiltonian
\begin{equation}
H_{\text{eff}}=U_{\text{rot}}(H_0+H_{\text{MW}})U_{\text{rot}}^\dagger-iU_{\text{rot}}\frac{dU_{\text{rot}}^\dagger}{dt}.
\end{equation}
By dumping the fast oscillating terms (the rotating wave approximation), we have
\begin{equation}
\begin{split}
H_{\text{eff}}&=A_3(\tau)\sigma_z \otimes \mathbf{I}+B_0(\tau)\mathbf{I}\otimes\sigma_z+B_3(\tau)\sigma_z\otimes\sigma_z\\
&+\Omega_1(t)\cos(\phi_1)\sigma_x\otimes |\uparrow_n\rangle\langle\uparrow_n|\\&+\Omega_1(t)\sin(-\phi_1)\sigma_y\otimes |\uparrow_n\rangle\langle\uparrow_n|\\
&+\Omega_2(t)\cos(\phi_2)\sigma_x\otimes |\downarrow_n\rangle\langle\downarrow_n|\\&+\Omega_2(t)\sin(-\phi_2)\sigma_y\otimes |\downarrow_n\rangle\langle\downarrow_n|. \label{RWAEq}
\end{split}
\end{equation}

\begin{figure}[h]
	\centering
	\includegraphics[width=1\linewidth] {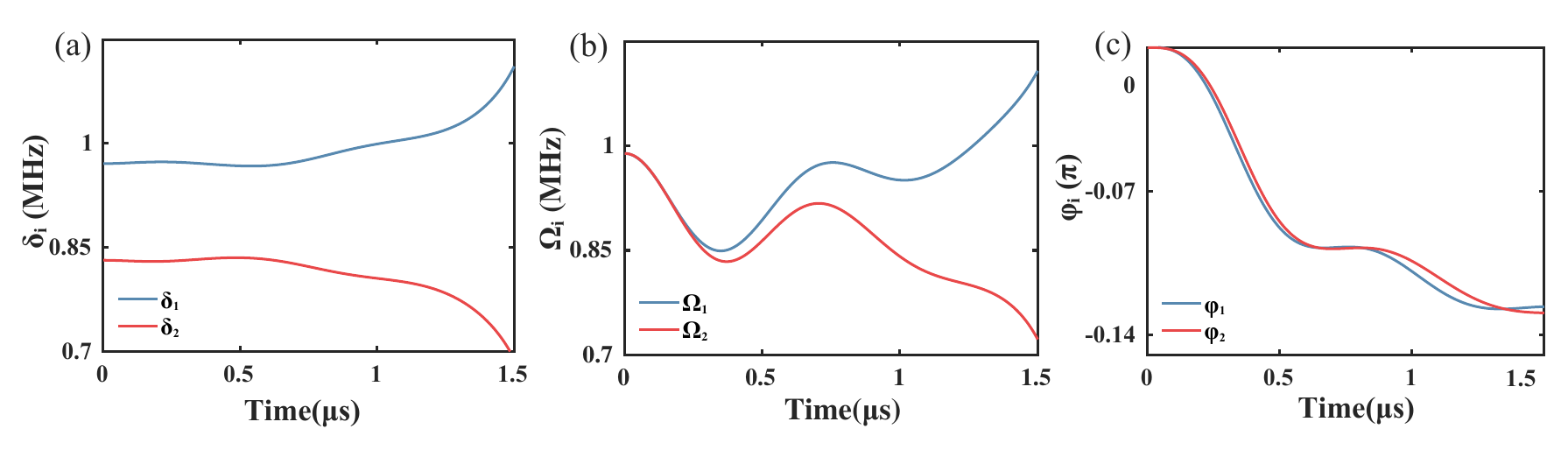}
	\caption{\textbf{Microwave parameters for the experiment}.
		The microwave frequency detuning, strength and phase are shown in \textbf{a}, \textbf{b} and \textbf{c} respectively. Here, the detuning means the difference between the MW frequency and the resonant frequency [i.e. $\delta_1(t)=\omega_1(t)-\omega_{\uparrow}$ and $\delta_2(t)=\omega_2(t)-\omega_{\downarrow}$]. 
	}
	\label{exp_parameter}
\end{figure}

Comparing Eq. \ref{RWAEq} with Eq. \ref{hsa}, we obtain the following experimental parameters: 
\begin{equation}
\left\{\begin{aligned}
\omega_1(t)&=\omega_{\uparrow}+2*(A_3(t)+B_3(t)),\\
\omega_2(t)&=\omega_{\downarrow}+2*(A_3(t)-B_3(t)),\\
\Omega_1(t)&=\frac{1}{\pi}\sqrt{(A_1(t)+B_1(t))^2+(A_2(t)+B_2(t))^2},\\
\Omega_2(t)&=\frac{1}{\pi}\sqrt{(A_1(t)-B_1(t))^2+(A_2(t)-B_2(t))^2},\\
\phi_1(t)&=-{\rm arctan2}(A_2(t)+B_2(t),A_1(t)+B_2(t)),\\
\phi_2(t)&=-{\rm arctan2}(A_2(t)-B_2(t),A_1(t)-B_2(t)).\\
\end{aligned}
\right.
\end{equation}
In Fig. \ref{exp_parameter}, we plot the time-dependence of the MW  detuning ($\delta_i$), amplitude ($\Omega_i$),  and phase ($\phi_i$).  

\section{Rotation of the Hamiltonian}
As mentioned in the main text, instead of measuring the $\langle \sigma_x\rangle$ directly, we rotate the whole system with  $U=\frac{1}{\sqrt{2}}\begin{pmatrix} 1&-1 \\1&1 \end{pmatrix}$ and then measure $\langle \sigma_z\rangle$ instead. In the following, we prove that these two methods are equivalent.
Let $|\psi\rangle$ be one of the eigen vector of Hamiltonian $H_e$, such that $H_e|\psi\rangle=\lambda|\psi\rangle$.
Then we have:
\begin{eqnarray}
\langle \psi|\sigma_x|\psi\rangle=\langle \psi|UU^\dagger\sigma_xUU^\dagger|\psi\rangle=\langle \widetilde{\psi}|\sigma_z|\widetilde{\psi}\rangle,
\end{eqnarray}
where $|\widetilde{\psi}\rangle=U^\dagger|\psi\rangle$ is the eigen vector of $\widetilde{H_e}$ with the same eigenvalue $\lambda$ due to the following equation
$$\widetilde{H_e}|\widetilde{\psi}\rangle=U^\dagger H_e UU^\dagger|\psi\rangle=U^\dagger \lambda|\psi\rangle=\lambda |\widetilde{\psi}\rangle.$$

\section{Relations between expectation values for different eigenstates}
For a Hermitian Hamiltonian, the eigenstates with different eigenvalues are orthogonal to each other. However, this is not necessarily true for non-Hermitian Hamiltonians. In general, the non-Hermitian SSH Hamiltonian considered in this work has two right eigenvectors ($|R_1\rangle$ and $|R_2\rangle$) and two left eigenvectors ($|L_1\rangle$ and $|L_2\rangle$). The Hamiltonian reads:
\begin{equation}
H(k)=\gamma[ (v+r\cos k) \sigma_x +(r\sin k+i/2)\sigma_z ],
\end{equation}
The right eigenstates are:
\begin{equation}
|R_1\rangle=\begin{pmatrix}
\cos \frac{\theta}{2}\\
-\sin \frac{\theta}{2}
\end{pmatrix},
|R_2\rangle=\begin{pmatrix}
\sin \frac{\theta}{2}\\
\cos \frac{\theta}{2}
\end{pmatrix}.
\end{equation}\\
And the left eigenstates are:
\begin{equation}
|L_1\rangle=\begin{pmatrix}
\cos \frac{\theta^*}{2}\\
-\sin \frac{\theta^*}{2}
\end{pmatrix},
|L_2\rangle=\begin{pmatrix}
\sin \frac{\theta^*}{2}\\
\cos \frac{\theta^*}{2}
\end{pmatrix}.
\end{equation}\\
where $\tan \theta=-(v+r\cos k)/(r \sin k+ i/2)$.
Now we calculate the expectation values of $\sigma_{x,y,z}$ for these eigenstates as follows:
\begin{scriptsize}
\begin{eqnarray}
\langle R_1|R_1\rangle&=&\langle R_2|R_2\rangle=\langle L_1|L_1\rangle=\langle L_2|L_2\rangle=\cos\left(\frac{\theta-\theta^*}{2}\right) \label{renormalize}\\
\frac{\langle R_1|\sigma_x|R_1 \rangle}{\langle R_1|R_1\rangle}&=&-\frac{\langle R_2|\sigma_x|R_2\rangle}{\langle R_2|R_2\rangle}=\frac{\langle L_1|\sigma_x|L_1\rangle}{\langle L_1|L_1\rangle}\\&=&-\frac{\langle L_2|\sigma_x|L_2\rangle}{\langle L_2|L_2\rangle}=-\frac{\sin(\frac{\theta+\theta^*}{2})}{\cos(\frac{\theta-\theta^*}{2})}\label{sigx}\\
\frac{\langle R_1|\sigma_y|R_1\rangle}{\langle R_1|R_1\rangle}&=&\frac{\langle R_2|\sigma_y|R_2\rangle}{\langle R_2|R_2\rangle}=-\frac{\langle L_1|\sigma_y|L_1\rangle}{\langle L_1|L_1\rangle}\\&=&-\frac{\langle L_2|\sigma_y|L_2\rangle}{\langle L_2|L_2\rangle}=i\frac{\sin(\frac{\theta-\theta^*}{2})}{\cos(\frac{\theta-\theta^*}{2})}\label{sigy}\\
\frac{\langle R_1|\sigma_z|R_1\rangle}{\langle R_1|R_1\rangle}&=&-\frac{\langle R_2|\sigma_z|R_2\rangle}{\langle R_2|R_2\rangle}=\frac{\langle L_1|\sigma_z|L_1\rangle}{\langle L_1|L_1\rangle}\\&=&-\frac{\langle L_2|\sigma_z|L_2\rangle}{\langle L_2|L_2\rangle}=\frac{\cos(\frac{\theta+\theta^*}{2})}{\cos(\frac{\theta-\theta^*}{2})}\label{sigz}
\end{eqnarray}
\end{scriptsize}
In the experiment, the initial state of the electron spin will decay to one of these eigenstates, and state crossing may occur at certain momentum points. As a result, we may need the above relations to draw the whole trajectory of the eigenstates on the Bloch sphere. Moreover, later in this Supplementary Information, we need the left eigenstates to calculate the topological index. We can use these relations to reconstruct the states with our experimental data.

\section{Direct calculation of the topological index with experimental data}
In the experiment, we have chosen three groups of Hamiltonian parameters corresponding to three different topological regions, with winding number $w=0,\frac{1}{2},1$ respectively.  For each case, we discretize the first Brillouin zone and choose a list of  momentum points $k$ from 0 to $2\pi$.  We employ the dilation method to prepare the electron spin onto an eigenstate of the non-Hermitian Hamiltonian $H(k)$ and measure $\sigma_x$ and $\sigma_z$ with varying $k$.
The $k$ points we used in the experiment and the corresponding results are shown in Tables \ref{t1}, \ref{t2} and \ref{t3}. 

\begin{table*}
	\centering
	
	\begin{tabular}{>{\raggedleft}p{20 pt}|>{\raggedleft}p{50 pt}>{\raggedleft}p{50 pt}|>{\raggedleft}p{50 pt}>{\raggedleft\arraybackslash}p{50 pt}}
		\hline
		\multirow{2}{*}{k}&\multicolumn{2}{c}{Expectation value: $\langle\sigma_x\rangle$}&\multicolumn{2}{c}{Expectation value: $\langle\sigma_z\rangle$}\\
		& experiment&theory& experiment&theory\\
		\hline
		0$\pi$&-0.07(10)&0.000&0.31(3)&0.280\\
		0.1$\pi$&0.15(5)&0.236&0.41(4)&0.414\\
		0.2$\pi$&0.26(5)&0.297&0.47(5)&0.554\\
		0.4$\pi$&0.26(3)&0.266&0.74(3)&0.761\\
		0.6$\pi$&0.18(3)&0.164&0.87(3)&0.890\\
		0.8$\pi$&0.12(6)&0.065&0.95(3)&0.954\\
		1.2$\pi$&-0.12(6)&-0.065&0.92(12)&0.954\\
		1.4$\pi$&-0.15(6)&-0.164&0.84(3)&0.890\\
		1.6$\pi$&-0.26(3)&-0.266&0.74(6)&0.761\\
		1.8$\pi$&-0.30(4)&-0.297&0.55(5)&0.554\\
		1.9$\pi$&-0.19(6)&-0.236&0.38(6)&0.414\\
		\hline
	\end{tabular}
	\caption{\textbf{Experimental data for $v=0.3$ and $r=0.18$}. Theoretically, the winding number for this chosen parameter is $w=0$. The second (fourth) column shows the experimentally measured data of $\langle \sigma_x\rangle$ ($\langle \sigma_z\rangle$). The third (fifth) column shows the experimental prediction. The experimental data agrees well with the theoretical prediction.}
	\label{t1}
\end{table*}

\begin{table*}
	\centering
	
	\begin{tabular}{>{\raggedleft}p{20 pt}|>{\raggedleft}p{50 pt}>{\raggedleft}p{50 pt}|>{\raggedleft}p{50 pt}>{\raggedleft\arraybackslash}p{50 pt}}
		\hline
		\multirow{2}{*}{k}&\multicolumn{2}{c}{Expectation value: $\langle\sigma_x\rangle$}&\multicolumn{2}{c}{Expectation value: $\langle\sigma_z\rangle$}\\
		& experiment&theory& experiment&theory\\
		\hline
		0.1$\pi$&0.49(3)&0.547&0.24(4)&0.268\\
		0.2$\pi$&0.48(4)&0.517&0.41(3)&0.482\\
		0.3$\pi$&0.51(3)&0.457&0.65(4)&0.647\\
		0.4$\pi$&0.42(3)&0.375&0.76(3)&0.775\\
		0.5$\pi$&0.29(3)&0.280&0.86(3)&0.870\\
		0.6$\pi$&0.09(3)&0.184&0.95(3)&0.935\\
		0.7$\pi$&0.09(4)&0.099&0.96(3)&0.975\\
		1.0$\pi$&0.04(8)&0.000&1.00(1)&1.000\\
		1.3$\pi$&-0.10(4)&-0.099&0.92(3)&0.975\\
		1.4$\pi$&-0.18(5)&-0.184&0.87(3)&0.935\\
		1.5$\pi$&-0.24(4)&-0.280&0.77(4)&0.870\\
		1.6$\pi$&-0.32(3)&-0.375&0.72(3)&0.775\\
		1.7$\pi$&-0.37(4)&-0.457&0.66(4)&0.647\\
		1.8$\pi$&-0.42(4)&-0.517&0.46(4)&0.482\\
		1.9$\pi$&-0.47(4)&-0.547&0.20(4)&0.268\\
		\hline
	\end{tabular}
	\caption{\textbf{Experimental data for v=0.3, r=0.3}. Theoretically, the winding number for this chosen parameter is $w=\frac{1}{2}$.}
	\label{t2}
\end{table*}

\begin{table*}
	\centering
	
	\begin{tabular}{>{\raggedleft}p{20 pt}|>{\raggedleft}p{50 pt}>{\raggedleft}p{50 pt}|>{\raggedleft}p{50 pt}>{\raggedleft\arraybackslash}p{50 pt}}
		\hline
		\multirow{2}{*}{k}&\multicolumn{2}{c}{Expectation value: $\langle\sigma_x\rangle$}&\multicolumn{2}{c}{Expectation value: $\langle\sigma_z\rangle$}\\
		& experiment&theory& experiment&theory\\
		\hline
		0.1$\pi$&0.80(3)&0.891&0.27(3)&0.259\\
		0.2$\pi$&0.76(3)&0.798&0.47(3)&0.499\\
		0.3$\pi$&0.59(4)&0.650&0.68(3)&0.705\\
		0.4$\pi$&0.34(3)&0.459&0.90(4)&0.864\\
		0.5$\pi$&0.25(6)&0.235&0.93(3)&0.965\\
		0.6$\pi$&0.09(7)&-0.007&0.91(3)&1.000\\
		0.7$\pi$&-0.26(5)&-0.252&0.85(3)&0.956\\
		0.8$\pi$&-0.48(3)&-0.477&0.68(3)&0.815\\
		0.9$\pi$&-0.54(4)&-0.644&0.42(3)&0.530\\
		1.1$\pi$&-0.51(4)&-0.644&-0.38(4)&-0.530\\
		1.2$\pi$&-0.46(5)&-0.477&-0.75(3)&-0.815\\
		1.3$\pi$&-0.26(5)&-0.252&-0.91(4)&-0.956\\
		1.4$\pi$&-0.03(5)&-0.007&-0.90(4)&-1.000\\
		1.5$\pi$&0.10(3)&0.235&-0.85(3)&-0.965\\
		1.6$\pi$&0.38(4)&0.459&-0.76(3)&-0.864\\
		1.7$\pi$&0.58(4)&0.650&-0.64(4)&-0.705\\
		1.8$\pi$&0.71(4)&0.798&-0.46(4)&-0.499\\
		1.9$\pi$&0.84(3)&0.891&-0.21(5)&-0.259\\
		\hline
	\end{tabular}
	\caption{\textbf{Experimental data for v=0.3, r=1}. Theoretically, the winding number for this chosen parameter is $w=1$.  From $k=1.1\pi$ to $k=1.9\pi$, we have changed the sign of the experimentally measured $\langle\sigma_{x,z}\rangle$ due to the eigenstate crossing discussed in the main text.}
	\label{t3}
\end{table*}

Now, we show how to calculate the winding number by using our experimentally data.   For a non-Hermitian Hamiltonian, the winding number is defined as \cite{lieu2018topological}
\begin{equation}
w=\frac{i}{\pi}\int_{BZ} \langle L^n_k|\frac{\partial}{\partial k}|R^n_k\rangle dk \label{topo_index},
\end{equation} 
where $|L^n_k \rangle$ and $|R^n_k\rangle$ are the left and right eigenvectors respectively as we mentioned before, $n$ labels the band index. Ref. \cite{fukui2005chern} proposes a more efficient way(Eq. \ref{DiscreteTopIndex}) to calculate the complex winding number for the discretized Brillouin zone. This allows us to calculate the winding number with our discretized experimental data.
\begin{equation}
\left\{\begin{aligned}
A_i=\ln \frac{\langle L^n_{k_{i+1}}|R^n_{k_i}\rangle}{\langle L^n_{k_i}|R^n_{k_i}\rangle}, \\
w=\frac{1}{\pi}\Sigma_i  \text{Im}(A_i).
\end{aligned}
\right. \label{DiscreteTopIndex}
\end{equation}
For the $r=0.3$ case, the Hamiltonian encircles only one exceptional point and its eigenstate is $4\pi$ periodic for the momentum $k$. Thus, $k$ must sweep through $4\pi$ to close the trajectory. In our experiment, we do not sweep $k$ for the region from $2\pi$ to $4\pi$. But, we can use the relations in Eqs. (\ref{sigx}-\ref{sigz}) to deduce $\langle \sigma_{x,y,z}\rangle$ for this region. In addition, all the left eigenstates can be obtained from those relations as well. In Table I in the main text, we show the winding numbers calculated directly by using our experimental data. From this table, it is clear that the calculated winding number matches its theoretical predictions excellently. For the $r=0.3$ case, we find that the winding number is 1 if we integrate $k$ from $0$ to $4\pi$ and conclude that $w=\frac{1}{2}$ if $k$ only sweeps from $0$ to $2\pi$, which is consistent with Ref.\cite{Lee,Viyuela2016}

\bibliographystyle{apsrev4-1-title}
\bibliography{Reference}

\end{document}